\documentclass[12pt,preprint]{aastex}

\slugcomment{Accepted to ApJ}

\shorttitle{Evaporation of grain-surface species by shock waves in proto-planetary disk}
\shortauthors{Aota, Inoue $\&$ Aikawa}

\begin{document}


\title{Evaporation of grain-surface species by shock waves in protoplanetary disk}


\author{Takuhiro Aota\altaffilmark{1}, Tsuyoshi Inoue\altaffilmark{2}, AND  Yuri Aikawa\altaffilmark{1} }

\altaffiltext{1}{Department of Earth and Planetary Sciences, Kobe University, Kobe 657-8501,Japan}
\altaffiltext{2}{National Astronomical Observatory of Japan}







\begin{abstract}
Recent ALMA (Atacama Large Millimeter/submillimeter Array) observations of young protostellar objects detected warm SO emission,
which could be associated with a forming protostellar disk.
In order to investigate if such warm gas can be produced by accretion shock onto the forming disk, 
we calculate the sputtering and thermal desorption of various grain surface species
in one dimensional shock waves.
We find that thermal desorption is much more efficient than the sputtering in the post-shock region.
While H$_{2}$O can be thermally desorbed, if the accretion velocity is larger than 8 km s$^{-1}$ with the pre-shock gas number density of 10$^{9}$ cm$^{-3}$,
SO is desorbed, if the accretion velocity $\gtrsim$ 2 km s$^{-1}$ and $\gtrsim$ 4km s$^{-1}$, with the pre-shock density of 10$^{9}$ cm$^{-3}$ and 10$^{8}$ cm$^{-3}$, respectively.
We also find that the column density of hydrogen nuclei in warm post-shock gas is $N_{ {\rm warm} } \sim 10^{21}$ cm$^{-2}$.

\end{abstract}





\section{INTRODUCTION}
Volatile elements, such as carbon, nitrogen, oxygen, sulfur, and phosphorus, are important ingredients
in the atmosphere and biosphere of planets.
Since molecules carrying these elements are readily found in the gas phase and/or ice mantles on dust grains
in molecular clouds,
the question is if and how they are delivered to protoplanetary disks and then to planets.
In our solar system, the relative abundances of volatiles to water in comets are observed to be similar to those in interstellar ice,
which suggests that interstellar ice could have been the raw material of the Solar system, at least in comet forming regions (\citealt{mc11}).

In recent years, theoretical models of the chemistry in protostellar cores and in forming disks have been constructed by several groups.
As the infall material enters the central regions of protostellar cores, complex organic molecules can be formed via grain-surface reactions on warm grains (\citealt{gh06, hv09}).
In spherical symmetric models of a protostellar core, molecules are sublimated at their sublimation temperatures, forming an onion-like distribution of molecular abundances in the envelope
(e.g. \citealt{rc03, awgh08, awhgh12, vvdd09, vdv11}).
In axisymmetric core models, which can form disks, the temperature distribution is not spherically symmetric, and the chemical composition of an infalling fluid parcel depends on its trajectory;
if the fluid parcel experiences only low temperatures, a significant amount of interstellar ice can survive to reach the disk (\citealt{vvdd09, vdv11, fanhw13, hwchg13}).

Since the infalling flow is usually supersonic, accretion onto protoplanetary disks is accompanied by a shock wave.
The kinetic energy of the bulk flow is converted to thermal energy.
While the gas is decelerated by the enhanced pressure at the shock front, the dust grains are not, and they fall into the disk.
The dust grains are heated by gas drag and energy transfer from shock heated gas.
The ice mantles are thermally sublimated (e.g. \citealt{hh91, lerh91, nh94}) and/or sputtered by collisions with gas particles (e.g. \citealt{b78,ds79,jcmh08}).
\cite{lerh91}  investigated sublimation of volatiles by the drag heating in the primitive solar nebula.
They showed that 90 $\%$ of water ice is sublimated by the drag heating at the radius of 30 AU.
The water vapor then re-condenses onto the grain surfaces, because the solar nebula is cold.
\cite{nh94}, on the other hand, performed more general calculations of dense molecular shocks
with pre-shock density $10^{7.5}-10^{12}$ cm$^{-3}$ and velocity  $5\le v \le 100$ km s$^{-1}$.
They obtained grain vaporization criteria for various grain materials such as water ice, organic refractory materials, silicate, and iron.

Observational studies on chemistry in these protostellar cores and disk forming stages have made
significant progress in the last decades.
Various complex organic molecules (COMs) are detected in hot corino sources, such as IRAS16293-2422
(\citealt{caselli12} and references therein),
while unsaturated carbon chains such as C$_4$H and c-C$_3$H$_2$ are found in L1527 and IRAS15398-3359 (\citealt{sakai08,sakai09,sakai13}).
These emission lines are generally considered to originate in the warm envelope heated by the protostar(s), rather than
accretion shock. \cite{sakai10} found that c-C$_3$H$_2$ abundance
is enhanced in the protostellar envelope at the radius of $500-1000$ AU, where the dust temperature reaches the
sublimation temperature of CH$_4$. Abundances of COMs in hot corinos are derived by assuming a spherical
envelope with radial gradients of density and temperature, although high-resolution observations have found
more complex physical structures, such as infall, outflow and binarity (\citealt[e.g.][]{jorgensen07,takakuwa07,pmf12, zlr13}).

Recently, \cite{ssh14} detected SO gas towards L1527.
The position velocity diagram (P-V diagram) indicates that SO is located and rotating in a ring region
at the centrifugal barrier of radius $\sim 100$ AU.
The temperature ($> 60$ K) derived from the LVG analysis of SO lines is higher than
the estimated sublimation temperature of SO ($\sim 50$ K), which indicates that SO is desorbed to the gas phase at this radius.
According to the theoretical models of protostellar cores (\citealt{awhgh12}), SO ice could indeed be abundant in
infalling material. \cite{ssh14} considered three possible desorption mechanisms: accretion shock in front of the
centrifugal barrier, outflow shock, and protostellar heating. The compactness of the SO emission and the P-V diagram
indicate that the outflow shock could not be the major cause of the enhancement of SO.
With regards to the protostellar heating, \cite{thc13} constructed a detailed model of L1527 by
solving radiation transfer and considering existing observational data in the infrared and radio wavelengths.
The derived temperature at the radius of 100 AU is about 30 K, which is in agreement with the excitation
temperature of c-C$_3$H$_2$ ($23-33$ K) in the infalling gas at $\sim 100$ AU \citep{ssh14}.
Although contributions of the outflow shock and protostellar heating are not completely ruled out, the 
coincidence of the SO emitting region with the centrifugal barrier is most naturally explained
by SO desorption via accretion shock.
It should also be noted that the SO emission detected towards L1489 IRS shows a similar P-V diagram
to that of L1527 \citep{yen14}.

If the SO emission is really tracing the accretion shock onto the forming disk, it will have a significant impact on the studies
of the physics and chemistry of disk formation.
Since the accretion flux is largest around the temporal centrifugal radius (\citealt{h98}),
SO emission could be a useful probe of disk formation.
Chemical composition of the infall material might be altered at the accretion shock.
Although \cite{lerh91} predicted that the vapor simply re-condenses in the disk, they did not include chemical 
reactions in their model.
It is thus important to confirm if the warm SO really originates in the accretion shock.
On the observational side, high resolution observations of the emission lines are desirable to distinguish which
desorption mechanism, accretion shock, outflow shock, or protostellar heating, is responsible for the SO emission
around L1527. On the theoretical side, it is useful to investigate the critical condition for
SO and other volatiles to be desorbed to the gas phase in the accretion shock.

In this work, we investigate the desorption of volatile species in the accretion shock onto protoplanetary disks.
Our goals are twofold: to investigate (1) if the SO emission detected toward L1527 is consistent with an 
accretion shock model and (2) what other molecules can be detected in the accretion shock.
We calculate the physical structure of a one-dimensional J-shock with pre-shock gas density and velocity relevant to disk formation.
Then we evaluate the drag heating of grains and a fraction of volatile molecules desorbed to the gas phase via thermal desorption and sputtering.
While \cite{nh94}  investigated a wide range of pre-shock velocity, we concentrate on low pre-shock velocities $\lesssim 10$ km s$^{-1}$ 
and investigate desorption of various volatile molecules which were not included in \cite{nh94}, in addition to water.
This paper is organized as follows.
In \S 2, we describe our physical and chemical models. In \S 3, we show the resultant shock structure and the fraction of
volatiles sublimated as a function of pre-shock density and velocity.
We also evaluate the column density of warm gas in the post-shock region.
In \S 4, we compare our model results with the observation of L1527.
Finally, we summarize our results in \S 5.

\section{MODEL}
\begin{figure}
\epsscale{.50}
\begin{center}
\fbox{
\plotone{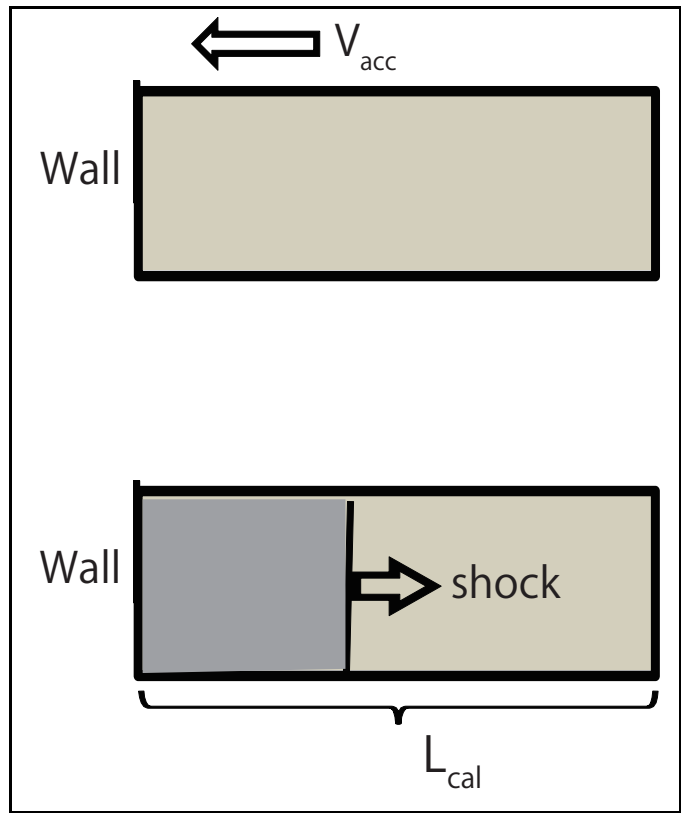}
}
\end{center}
\caption{Schematic view of our 1D shock model}
\label{cal-picture}
\end{figure}

We investigate the accretion shock onto protoplanetary disk using 1D plane-parallel model.
Figure \ref{cal-picture} schematically shows the configuration of our model.
We consider a collision of gas flow with a rigid wall.
This configuration also corresponds to the head-on collision between two gas flows with the same density.
Our calculations consist of two steps.
Firstly, we calculate the physical structure of the shock, i.e. density, temperature, and velocity of the gas.
Secondly, we calculate the dust velocity, temperature, and the desorption of ice mantles in the post shock region.
We explain the basic equations in the following subsections.

\subsection{Structure of 1D Shock}
We perform one dimensional hydrodynamic simulations.
We solve the conservation of mass, momentum, and energy:

\[ \partial_{t}U(t,x)+\partial_{x}F_{x}=S  \]
\[ U=(\rho,\rho v_{x},E) \]
\[ F_{x}= \left( \begin{array}{ccc} \rho v_{x} \\ 
\rho v_{x}^{2} + p \\  
 (E+p) v_{x} \end{array} \right) \]
\[ S= \left( \begin{array}{ccc} 0 \\
0 \\
- \rho\,\Lambda \end{array} \right) \]
\[ E=\frac{p}{\gamma - 1} + \frac{\rho v^{2}_{x} }{2} \quad ,\]
where $\rho, v_{x}, p$, $\gamma$, and $\Lambda$ are the gas mass density, velocity, thermal pressure, ratio of specific heats, and
cooling rate per unit mass, respectively.
External heating, i.e. the protostellar heating, is not explicitly included in our model. It is, however, implicitly
taken into account by setting the initial and minimum gas temperature (see below).

We use an operator-splitting technique to solve these equations, which are split into two parts: (1) ideal hydrodynamics,
and (2) cooling (e.g., \citealt{ii08}).
The former is calculated by the second-order Godunov method with Lagrangian coordinates \citep{v79}.
We solve the exact Riemann problem iteratively at each grid cell interface to calculate numerical fluxes,
and determine the position of grid cell interface in the next time step.
After solving ideal hydrodynamics, the energy equation
\[ \frac{\partial E}{\partial t} = - \rho\, \Lambda  \]
is solved by the second-order explicit method.

In this work, we consider high density gas ($n_{{\rm H}} \ge 10^{6}$ cm$^{-3}$) accreting onto protoplanetary disk.
At such high densities, gas is cooled by gas-dust collisional cooling.
Referring to \cite{hh91}, the cooling rate per unit volume is
\begin{equation}
\label{gas-cooling}
\rho\, \Lambda = \frac{1}{2} \pi^{0.5} a_{{\rm dust}}^{2} n_{{\rm dust}} \rho_{{\rm gas}} (T_{{\rm gas}}-T_{{\rm dust}}) \frac{\gamma + 1}{\gamma - 1}  \left( \frac{ {2k_{{\rm bol}} } }{m_{{\rm gas}}}  \right)^{1.5} T_{{\rm gas}}^{0.5}, 
\end{equation}
where $a_{{\rm dust}}$, $ n_{{\rm dust}} $, $k_{{\rm bol}}$, and $m_{{\rm gas}}$ are grain radius, number density of dust grain, Boltzmann constant, and mean mass of gas particle, respectively.
In this work, we set $a_{{\rm dust}}$ = 0.1 $\mu$m, which is a typical grain radius in interstellar clouds (\citealt{t05}), 
and the mean mass of gas is set to be $m_{ {\rm gas} } = 1.67 \times 10^{-24} \left( \frac{ 2 n_{ {\rm H}_{2} } + 4  n_{ {\rm He} } }{ n_{ {\rm H}_{2}} + n_{ {\rm He} } } \right) \sim 3.9 \times 10^{-24}$ g.
$T_{ {\rm dust} }$ is determined by the balance between heating and cooling: gas-grain energy transfer and cooling by thermal emission of dust (see section 2.2).
At this stage, we neglect the relative velocity between gas and dust; the heating rate of dust grains is given by equation (\ref{gas-cooling}).
The dust temperature derived here is slightly different from the value obtained in section 2.2., where we take into account the gas-dust drag.
But the difference is so small that it does not affect the gas temperature.
We neglect other cooling mechanisms such as CO rotational cooling, which are less efficient than gas-dust cooling at $n_{ {\rm H} }$ $\gtrsim$ 10$^{7}$ cm$^{-3}$.

The time step of the integration is set to be small enough to satisfy the Courant-Friedrichs-Lewy (CFL) condition: $0.03 \times \tau_{cool}$, where $\tau_{cool}$ is the cooling time.
In this work, we set the minimum and initial gas temperature to be 20 K.
We explore the parameter space of pre-shock number density of hydrogen nuclei $n_{ {\rm H} } =10^{6}, 10^{7}, 10^{8}$  and $10^{9}$ cm$^{-3}$.
We adopt this temperature and these densities for pre-shock gas referring to the recent three dimensional SPH simulation of disk formation and evolution with radiation transfer (\citealt{tmi13}).
The assumed initial temperature is slightly lower than the envelope temperature of L1527 derived from
the observation of c-C$_3$H$_2$ lines ($23-33$ K) (\citealt{ssh14}) and from radiation transfer modeling ($\sim 30$ K)
(\citealt{thc13}). Our results do not significantly depend on the initial temperature, as long as we are focused
on molecules with sublimation temperatures higher than the initial temperature. 
The accretion velocity $V_{ {\rm acc}}$ ranges from 1 km s$^{-1}$ to 10 km s$^{-1}$, which corresponds to the free fall velocity at $r \gtrsim 15$AU around a Solar mass protostar.
These velocities are chosen considering the spatial resolution of ALMA observation.
The highest spatial resolution of line observation by ALMA is of order 0.1 arc second.
If we consider the nearest star forming regions such as Taurus, the spatial resolution is about 15 AU.
Our numerical domain varies from 1.0$\times 10^{-3}$ pc to 2.0$\times 10^{-6}$ pc depending on the model parameters (see Table \ref{parameter}).
The numerical domain is divided into 450 grid cells with equal intervals.
Although our hydrodynamics code is time-dependent, we use it to obtain the steady state structure of 1D shock;
we run our code until the flow around the shock front reaches the steady state.

\subsection{Dust Velocity and Temperature}
While the gas is decelerated by shock, dust particle flows through the gas, and heated by the gas-dust collision.
In the 1D steady state shock obtained in \S 2.1., we calculate dust velocity relative to gas.
The equation of motion of dust is
\[  \frac{4}{3} \pi \rho_{ {\rm mat}} a_{ {\rm dust} }^{3} \frac{dV_{ {\rm dust} }}{dt} = - \pi a_{ {\rm dust} }^{2} \frac{C_{D}}{2} \rho V_{ {\rm dust} }^{2} \]
(\citealt{hh91}).
We solve the equation of motion by the 4th order Runge-Kutta method.
The time step of the integration is set to be  $0.01 \times \tau_{\rm stop}$. The dust stopping time $\tau_{\rm stop}$ is
\[  \tau_{ {\rm stop} }   = \frac{ 4\rho_{ {\rm mat}} a_{ {\rm dust}  }  }{ 3 \rho V_{ {\rm dust}}  } , \]
where $\rho_{ {\rm mat}} $, $V_{ {\rm dust} }$, $C_{D}$ are grain material density, grain velocity relative to the gas flow, and the drag coefficient, respectively.
In this work, $\rho_{ {\rm mat}}$ is set to be 3.0 g cm$^{-3}$. The drag coefficient ($C_{D}$) is given by (\citealt{hh91})
\[ C_{D} = \frac{ 2 }{ 3s_{a} } \left(  \frac{ \pi T_{ {\rm dust} } }{ T_{ {\rm gas} } } \right)  + \frac{ 2s_{a}^{2} + 1 }{ s_{a}^{3} \sqrt{\pi} } {\rm exp}( -s_{a}^{2} ) + \frac{ 4s_{a}^{4} + 4s_{a}^{2} -1 }{ 2s_{a}^{4} } {\rm erf}(s_{a}) , \]
where $s_{a} = V_{ {\rm dust } } /V_{T}  $ is the ratio of the relative grain velocity ($ V_{ {\rm dust } } $) to the gas thermal velocity ($V_{T} = \sqrt{ 2k_{ {\rm bol}} T_{ {\rm gas } } / m_{ {\rm gas } } } $).

The dust temperature is determined by the balance between
gas-grain energy transfer rate per dust grain ($\Gamma_{d}$) and cooling rate by thermal emission of dust ($\Lambda_{d}$):
\begin{equation}
\label{dust-heating}
\Gamma_{d} = 4 \pi a_{ {\rm dust} }^{2} \rho_{ {\rm gas} } V_{ {\rm dust}} ( T_{ {\rm rec} } - T_{ {\rm dust}  } ) C_{H}
\end{equation}
\[ \Lambda_{d} = 4 \pi a_{ {\rm dust} }^{2} \epsilon_{em} \sigma_{SB} T_{ {\rm dust} }^{4},  \]
where  $\epsilon_{em}$ and $\sigma_{SB}$ are emission coefficient and Stefan Boltzmann constant, respectively.
We adopt the emission coefficient ($\epsilon_{em}$) of silicate model in \cite{dl84} .
The adiabatic  recovery temperature, $ T_{ {\rm rec} } $,  and the heat transfer function, $ C_{H} $, are 
\[  T_{ {\rm rec} }  = \frac{ T_{ {\rm gas}}  }{  \gamma + 1 }  \left( 2 \gamma + 2( \gamma -1 )s_{a}^{2} - \frac{  \gamma -1  }{  1/2 + s_{a}^{2} + \frac{ s_{a} }{ \sqrt{\pi}  {\rm erf}(s_{a}) } {\rm exp}( -s_{a}^{2} )  }  \right)   \]
\[  C_{H} = \frac{  \gamma + 1 }{  \gamma - 1 }  \frac{  k_{ {\rm bol} }  }{ 8 m_{{\rm gas}} s_{a}^{2}  }  \left(  \frac{ s_{a}  }{  \sqrt{\pi} }  {\rm exp}( -s_{a}^{2} ) + \left(  1/2 + s_{a}^{2}  \right) {\rm erf}(s_{a})  \right) \]
(\citealt{hh91}). In the limit of $V_{ {\rm dust} } = 0$,  the equation (\ref{dust-heating}) is equal to equation (\ref{gas-cooling}) (see section 2 of \citealt{hh91}).

\subsection{Sputtering and Thermal Desorption}
We calculate the desorption rate of grain surface species along the flow of dust particles,
using the gas and dust temperatures, gas number density, and the relative velocity of dust to gas obtained in the previous subsections.
We calculate the rates of sputtering and thermal desorption separately to investigate which is more effective to desorb grain surface species.
When a dust particle of radius $a_{{\rm dust}}$ is moving with the velocity $V_{{\rm dust}}$ relative to gas, 
the sputtering rate of grain surface species $i$ per unit volume is 
\begin{equation}
\label{sputtering}
\frac{dn_{i}}{dt} = \frac{ n_{i} }{  \displaystyle \sum_{j} n_{j}} \pi a_{ {\rm dust} }^{2} n_{p} n_{{\rm dust}}  \int dx x^{2} \left( \frac{  8 k_{ {\rm bol} } T_{ {\rm gas} }  }{  m_{i} \pi  }  \right)^{0.5}  \frac{1}{2s} \left( e^{  -(x-s)^{2} } - e^{ -(x+s)^{2} } \right) \langle Y(E=x^{2}k_{ {\rm bol} }T_{ {\rm gas} }) \rangle_{\theta}, \\
\end{equation}
where
\begin{eqnarray}
x^{2} &=& \frac{1}{2} m_{p} v^{2}/k_{ {\rm bol} }T_{ {\rm gas} } = E/k_{ {\rm bol} }T_{ {\rm gas} }, \nonumber \\
s^{2} &=& \frac{1}{2} m_{p} V_{{\rm dust}}^{2}/k_{{\rm bol}} T_{{\rm gas}}  \nonumber
\end{eqnarray}
(\citealt{ds79}, see also \citealt{jcmh08}).
In Eq. (3), $n_{i}$, $n_{p}$, and $n_{{\rm dust}}$ are
the number densities of dust surface species $i$, projectile species, and dust grain, respectively. 
The parameters $x^2$ and $s^2$ represent the projectile kinetic energy of the random motion and the bulk motion relative to a dust
grain, normalized by the thermal energy, respectively.
The denominator in the right-hand side of Eq. (3) is the summation of number density of dust surface species.
The integral term is the average gas velocity relative to dust (see appendix \ref{mean-vel}).
The parameter $\langle Y(E) \rangle_{\theta}$ is the angle-averaged sputtering yield
\[ <Y>_{\theta} = 2 Y(E,\theta=0) = 2 \times 8.3 \times 10^{-4} \frac{ ( \epsilon - \epsilon_{0} )^{2} }{ 1 + (\epsilon/30)^{ \frac{ 4 }{ 3 } }   } ,~~ \epsilon > \epsilon_{0}, \]
\[  \epsilon = \frac{  \eta x^{2}k_{bol}T_{ {\rm gas} } }{ E_{b} }  \]
\[  \epsilon_{0} = {\rm Max} [1,4\eta]  \]
\[  \eta = 4 \times \xi m_{p}M_{t}(m_{p} + M_{t} )^{-2},  \]
where $E_{b}$ and $M_{t}$ are the binding energy and mass of the target species $i$  (\citealt{ds79}).
Binding energies of molecules onto grain surfaces are adopted from \cite{gh06}.
$\xi$ is 0.8 for ice and 1.0 for atoms on grain surfaces. In this work, we restrict ourselves to desorption of ice, and thus set $\xi$ to be 0.8.
Basically, when the kinetic energy of a projectile is larger than $4E_{b}$, the projectile can sputter dust surface species (see e.g. \citealt{b78, ds79, d95}).
The projectiles are gas particles of H$_{2}$, He and CO.

The thermal desorption rate is given by
\[  \frac{ dn_{i} }{ dt }  =  n_{i} \nu_{i}  {\rm exp}(- \frac{ E_{b} }{  k_{ {\rm bol} }T_{ {\rm gas} } }  ) \]
(e.g. \citealt{t05}).
The characteristic vibrational frequency of the dust surface species {\it i} is
\[ \nu_{i} = \sqrt{ \frac{ 2 \sigma E_{b}  }{  m_{i} \pi^{2} }  } , \]
where $\sigma = 1.5 \times 10^{15}$ cm$^{-2}$ is the surface density of binding sites.

We adopt the initial abundances of dust surface species from Nomura \& Millar (2004) (see Table \ref{abund}).
In this work, we assume that gas-dust mass ratio is 100:1.
We also investigate desorption of molecules which are not in Table \ref{abund}, such as SO.
In cases like this, we set their initial abundances to be $5.0 \times 10^{-8}$ relative to hydrogen (i.e. 1.8$ \times 10^{-4}$ relative to water ice).
The fraction of the sputtered amount over its initial abundance however, does not depend on the initial abundance,
as long as the abundance of the dominant ice, H$_{2}$O, is fixed (see equation (\ref{sputtering})).
In the following, we show this fraction rather than the absolute abundance of the desorbed species.

\section{RESULTS}

\subsection{Sputtering}
\begin{figure}
\epsscale{0.80}
\plotone{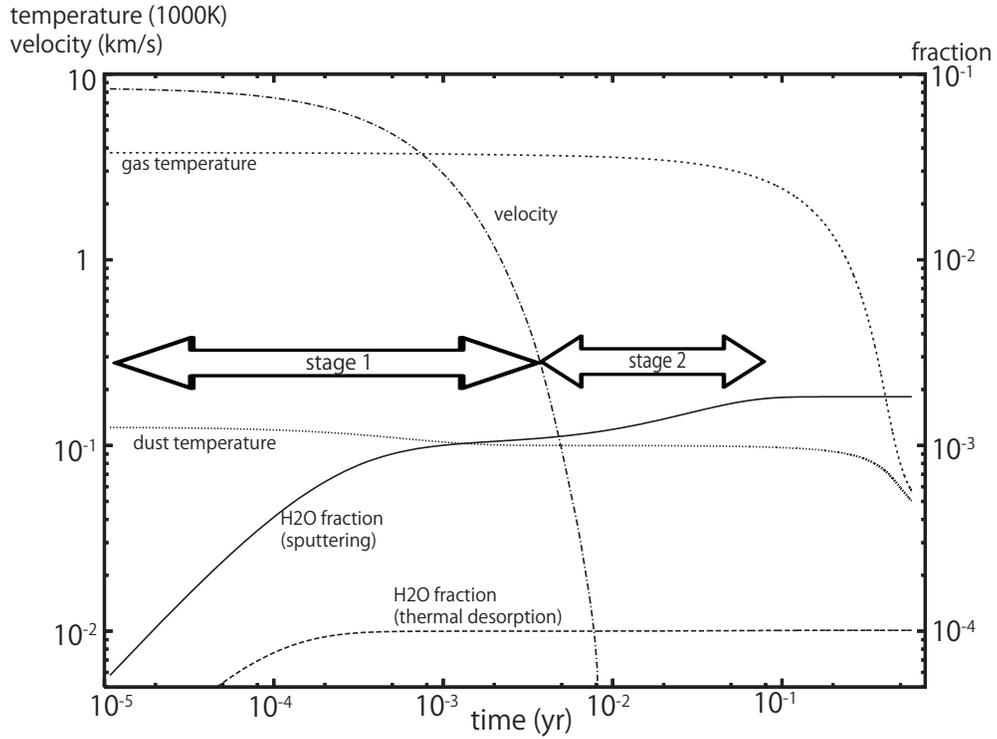}
\caption{ Temporal variations of the cumulative fraction of desorbed H$_{2}$O by sputtering and thermal desorption, 
gas and dust temperatures, and dust velocity relative to gas in the coordinate of a dust particle
in the model with pre-shock velocity $V_{ {\rm acc} } =$10 km s$^{-1}$ $V_{ {\rm acc} } =$10km s$^{-1}$ and density
$n_{ {\rm H} } = 10^{8}$ cm$^{-3}$.  }
\label{sputter-H2O-10km}
\end{figure}

We calculated the amount of molecules desorbed via sputtering by following a dust particle in the 1D shock model.
Figure \ref{sputter-H2O-10km} shows the results with pre-shock velocity $V_{ {\rm acc} }=10$ km s$^{-1}$ and density $n_{ {\rm H} }=10^{8}$ cm$^{-3}$.
The gas temperature, dust temperature, and the relative velocity of dust to gas are plotted as a function of time in the coordinate of the dust particle.
The moment when the dust particle passes through the shock front is set to $t=0$ in the figure.
The solid line depicts the cumulative fraction of sputtered H$_{2}$O along the flow over the initial H$_{2}$O ice abundance.
Although we do not plot gas density $n_{ {\rm H} }$ in this figure, it is about $6 \times 10^{8}$ cm$^{-3}$ in the post-shock gas (cf. \citealt{ll87})
until it increases sharply at $t \sim 0.1$ yr to keep the pressure constant as the gas cools.
We can see that the dust particle is sputtered by the friction between gas and dust until the dust particle is stopped by
the gas drag at $t\sim \tau_{ {\rm stop} } \sim 10^{-3}$ yr.
We call this phase stage 1. After stage 1, the dust particle is sputtered gradually by collisions with hot gas
until the gas is cooled in $t \sim \tau_{ {\rm cool} } \sim 0.1$ yr (stage 2).
When the gas temperature is high, the sputtering in stage 2 is effective in comparison with that in stage 1.

\begin{figure}
\epsscale{1.10}
\plotone{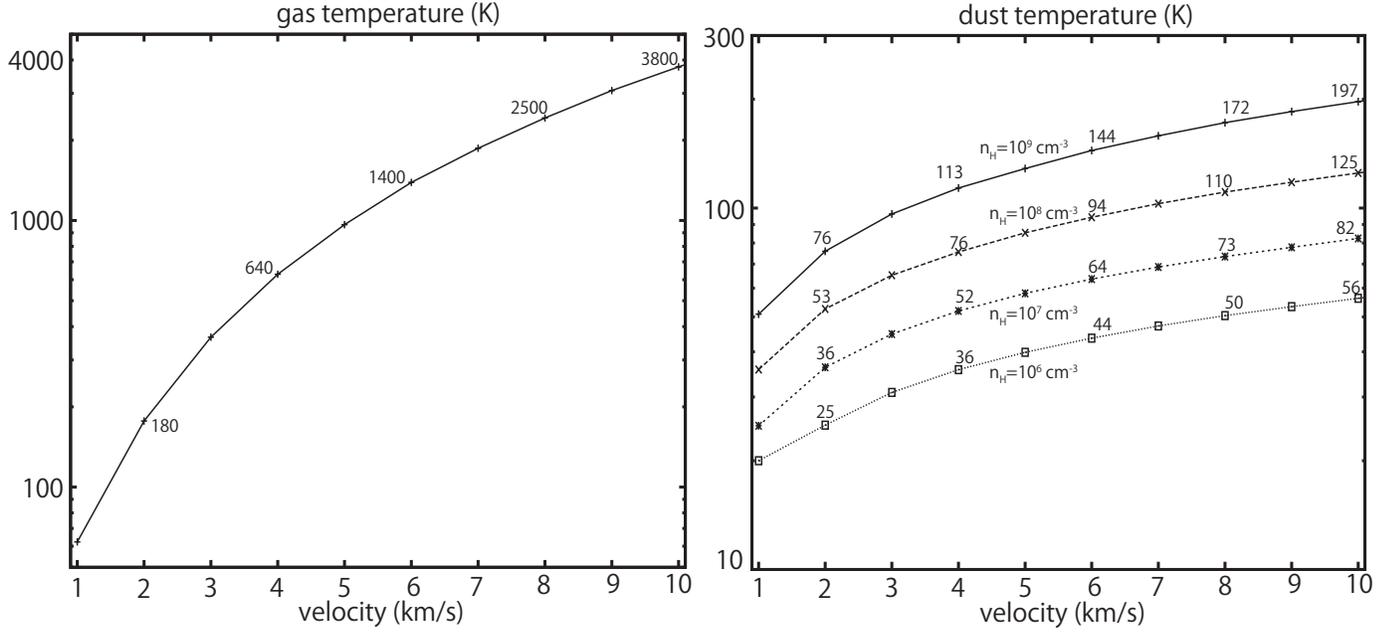}
\caption{The peak gas temperature (left) and peak dust temperature (right) as a function of pre-shock velocity.
The values of the vertical axis (gas or dust temperature) are labeled on some points.}
\label{gas-dust-temperature}
\end{figure}

Figure \ref{gas-dust-temperature} shows the peak temperatures of gas and dust as a function of pre-shock velocity
for assorted pre-shock gas densities.
The peak gas temperature is determined by the Rankine-Hugoniot relations (\citealt{ll87}), and does not depend on the
pre-shock density.
On the other hand, the peak dust temperature is determined by the balance between heating by collisions with gas
particles and cooling by thermal radiation (see Appendix \ref{d-tem}).
The dust temperature is higher, when the gas density is higher, because more frequent collisions with gas particles 
enhance the heating rate.

\begin{figure}
\epsscale{0.60}
\plotone{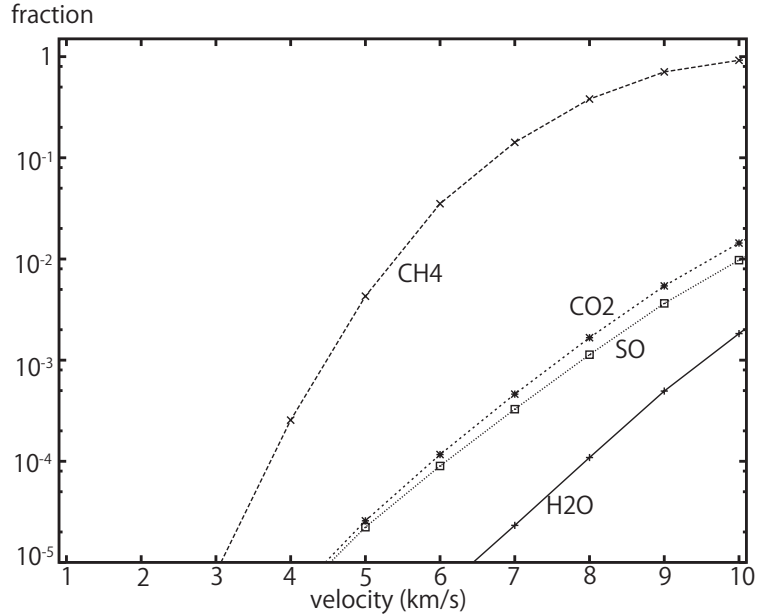}
\caption{The fraction of sputtered H$_{2}$O, SO, CO$_{2}$, and CH$_{4}$ to their initial ice abundances as a function of the pre-shock velocity.
The pre-shock density is fixed at $n_{ {\rm H} }=10^{8}$ cm$^{-3}$.}
\label{sputter-manysp-each-vel}
\end{figure}

We calculated the fraction of sputtered volatile molecules in various shock models.
Figure \ref{sputter-manysp-each-vel}  shows the fraction of sputtered H$_{2}$O, SO, CO$_{2}$, and CH$_{4}$
to their initial ice abundances as a function of the pre-shock velocity with a fixed pre-shock density of $n_{ {\rm H}} = 10^{8}$ cm$^{-3}$.
The binding energies ($E_{b}/k_{ {\rm bol} }$) of the molecules onto grain surfaces are set to be 5700 K for H$_{2}$O, 2600 K for SO, 2575 K for CO$_{2}$, and 1300 K for CH$_{4}$.
For example, the fraction of sputtered H$_{2}$O is only 0.2 $\%$ when the pre-shock velocity is 10 km s$^{-1}$.
The fraction is similar between SO and CO$_{2}$, because the binding energy of SO is close to that of CO$_{2}$.
We can see that a highly volatile species like CH$_{4}$ can be sputtered completely at 10 km s$^{-1}$.
We performed additional calculations covering higher pre-shock velocities to find that H$_{2}$O can be completely desorbed by the sputtering with $V_{ {\rm acc} }=19$ km s$^{-1}$;
10 $\%$ of H$_{2}$O is sputtered in stage 1, and the rest is sputtered in stage 2.
We speculate however, that our calculation overestimates the amount of sputtered water in such high velocity shocks;
the cooling by Ly-$\alpha$ and H$_{2}$ emission, which are not included in our model, could be efficient.
A higher pre-shock velocity $V_{ {\rm acc} } > 19$ km s$^{-1}$ is needed to completely desorb H$_2$O ice by sputtering.

\begin{figure}
\epsscale{0.60}
\plotone{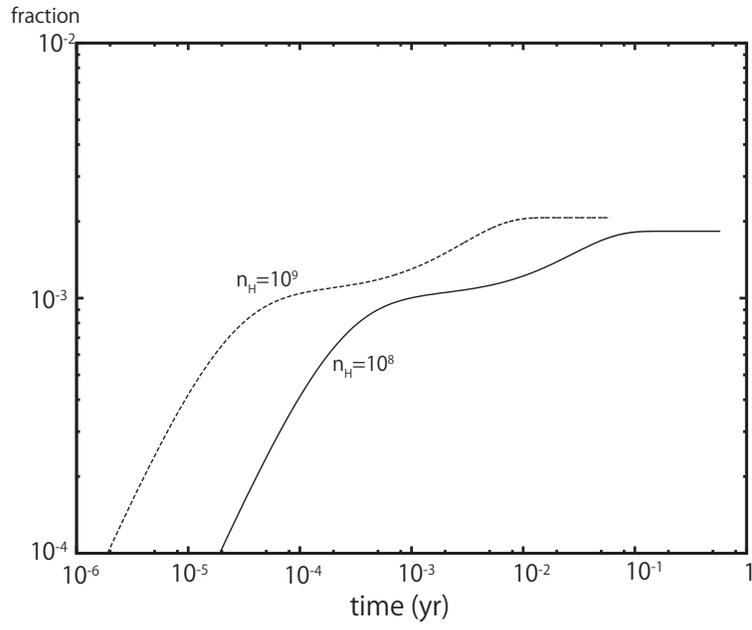}
\caption{Temporal variations of the fraction of sputtered H$_{2}$O when the pre-shock gas density is
$n_{ {\rm H}  } =10^{8}$ and $n_{ {\rm H}  }=10^{9}$ cm$^{-3}$.
The pre-shock velocity is fixed at $V_{ {\rm acc} } = 10$ km s$^{-1}$.}
\label{compare-sputtering-density}
\end{figure}

Figure \ref{compare-sputtering-density} shows the temporal variation of the cumulative fraction of sputtered H$_{2}$O in models with the
pre-shock densities $n_{ {\rm H} } =10^{8}$ and 10$^{9}$ cm$^{-3}$.
The pre-shock velocity is fixed at $V_{ {\rm acc} }=10$ km s$^{-1}$.
We can see that the total fraction of sputtered H$_{2}$O does not depend on pre-shock density,
although the sputtering completes earlier in the higher density model.
In stage 1, the amount of desorbed species $i$ ($N_{i}$) is proportional to the product of stopping time of dust ($\tau_{ {\rm stop}}$) and number density ($n_{ {\rm H} }$), i.e. $ N_{i} \propto n_{ {\rm H} }  \tau_{ {\rm stop} }$.
Since the stopping time is proportional to $1/n_{ {\rm H} }$, $N_{i}$ is independent of $n_{ {\rm H} }$.
In stage 2, the amount of desorbed species $i$ ($N_{i}$) is proportional to the product of cooling time of shocked gas ($\tau_{ {\rm cool} }$) and number density ($n_{ {\rm H} }$), i.e. $ N_{i} \propto n_{ {\rm H} }  \tau_{ {\rm cool} }$.
Since the cooling time is proportional to $1/n_{ {\rm H} }$, $N_{i}$ is again independent of $n_{ {\rm H} }$.

\subsection{Thermal Desorption}
The dashed line in Figure \ref{sputter-H2O-10km} shows the temporal variation of the cumulative fraction of thermally desorbed H$_{2}$O in the model with $n_{ {\rm H} }=10^{8}$ cm$^{-3}$ and $V_{{\rm acc}}$=10 km s$^{-1}$.
We can see that the thermal desorption occurs right after the dust particle passes the shock front.
The fraction of thermally desorbed H$_{2}$O is determined by the maximum dust temperature, which is achieved immediately behind the shock front.

\begin{figure}
\epsscale{1.10}
\plotone{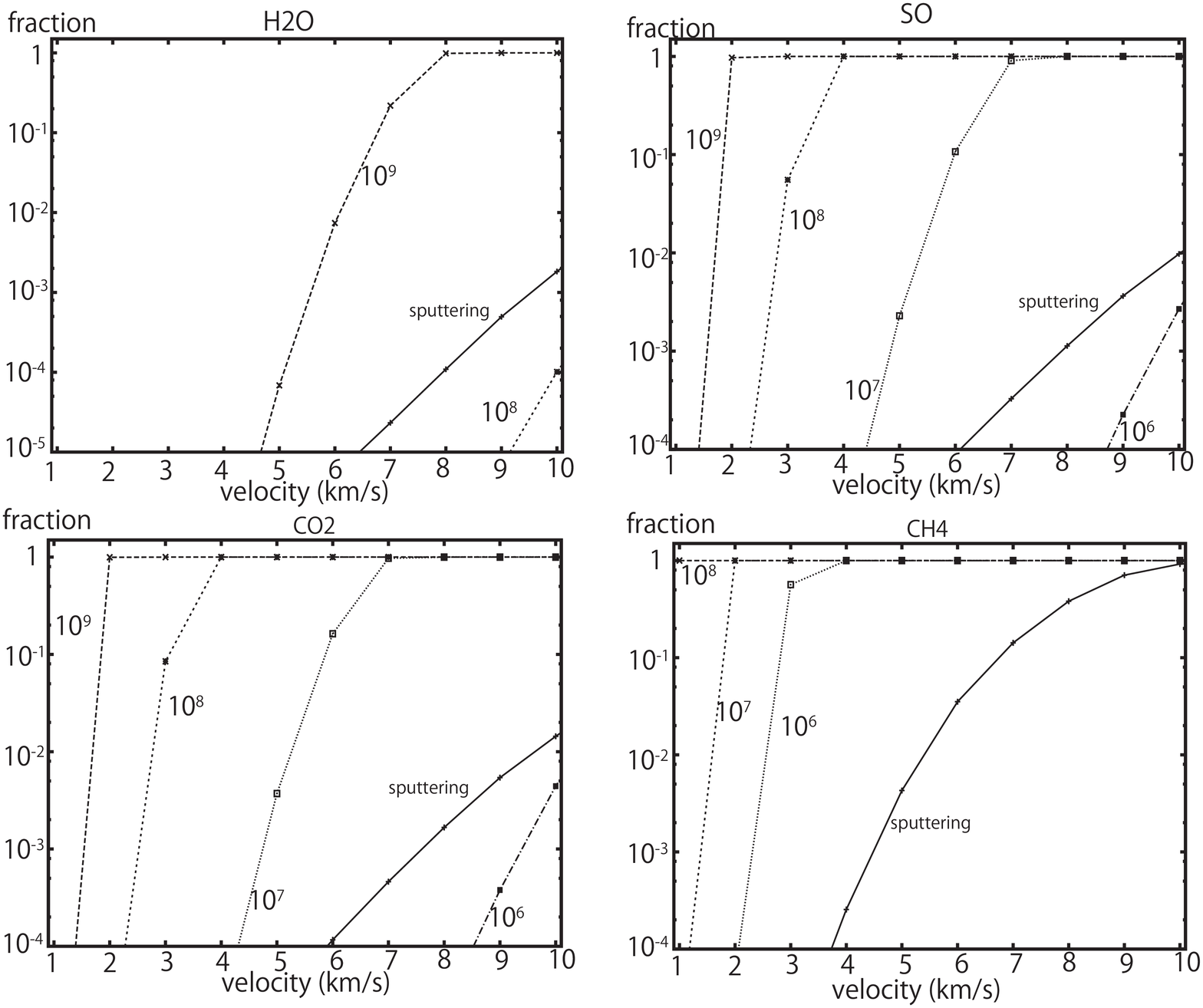}
\caption{The Fraction of thermally desorbed and sputtered H$_{2}$O, SO, CO$_{2}$, and CH$_{4}$ 
as a function of pre-shock velocity with the pre-shock gas density $n_{{\rm H}} = 10^{6}, 10^{7}, 10^{8}$, and $10^{9}$ cm$^{-3}$.}
\label{thermal-many-species}
\end{figure}

Figure \ref{thermal-many-species} shows the fraction of thermally desorbed H$_{2}$O, SO, CO$_{2}$, and CH$_{4}$ as functions of the pre-shock velocity for assorted pre-shock gas densities.
The fraction of sputtered molecules is shown with solid lines for comparison.
In most cases, thermal desorption is more effective than sputtering, because
the gas number density in our models is high enough to raise the dust temperature in the shock wave (see Appendix \ref{d-tem}).
Figure \ref{boundary} shows the critical  velocity above which the species is completely desorbed.
We can see that H$_{2}$O can be completely desorbed at $V_{ {\rm acc} } =$ 8 km s$^{-1}$, if the pre-shock density $n_{ {\rm H} }$ is $10^{9}$ cm$^{-3}$.
This threshold velocity is slightly lower than that obtained by \cite{nh94} ($\sim$10 km s$^{-1}$), because our binding energy of H$_{2}$O onto grain surfaces is lower than their value.
Except for this small shift in the critical velocity, our result is basically consistent with \cite{nh94}.
We can also see that SO and CO$_{2}$ can be desorbed at $V_{ {\rm acc} } \sim$ several km s$^{-1}$ when the preshock gas density is $n_{ {\rm H} } \ge 10^{7}$ cm$^{-3}$.
The most volatile species in our calculations, CH$_{4}$, can be desorbed at a lower gas density, i.e. $n_{ {\rm H} } = 10^{6}$ cm$^{-3}$.
In addition to the species discussed above, Figure \ref{boundary} shows the critical velocities for CH$_{3}$OH ($E_{b}/k_{ {\rm bol} }=$5534K)
HC$_{3}$N ($E_{b}/k_{ {\rm bol} }=$4580K), SO$_{2}$ ($E_{b}/k_{ {\rm bol} }=$3405K),
H$_{2}$CO ($E_{b}/k_{ {\rm bol} }=$2050K), and CS ($E_{b}/k_{ {\rm bol} }=$1900K).

\subsection{Column Density of Warm Gas}
The recent ALMA observation by \cite{ssh14} detected warm SO emission towards L1527.
The P-V diagram indicates that the warm SO exists in the ring region of radius $r \sim 100$ AU, which may be heated by the accretion shock from the protostellar envelope to the forming protoplanetary disk.
In this subsection, we calculate the column density of warm gas produced by the 1D shock.
Firstly, it is useful to analytically evaluate the column density of warm gas $N_{ {\rm warm} }$ behind the shock front.
When the gas temperature is much higher than the dust temperature, $N_{ {\rm warm} }$ can roughly be estimated
to be $n_{  {\rm pre} }  V_{  {\rm acc} }  \tau_{ {\rm cool} }$, 
where $n_{ {\rm pre} }$, $V_{ {\rm acc} }$, and $\tau_{ {\rm cool} }$ are the number density of pre-shock gas, accretion velocity, and cooling time scale, respectively.
The cooling time $\tau_{ {\rm cool} }$ is
\begin{equation}
\label{cool_t}
\tau_{cool} = p_{ {\rm post} }/( (\gamma -1) \Lambda ) \sim 6 \left(  \frac{ a_{ {\rm dust} } }{ 0.1 \mu m }  \right) \left(  \frac{ \rho_{ {\rm mat} } }{ 3.0 ~ {\rm g~ cm}^{-3} }  \right)  \left(  \frac{ R_{gd} }{ 100 }  \right)  \left(  \frac{ 100K }{ T_{ {\rm post} }  } \right)^{0.5}  \left(  \frac{ 10^{8} ~ {\rm cm}^{-3} }{ n_{ {\rm post} } }    \right)  ~~{\rm yr},
\end{equation}
where $p_{ {\rm post} }$, $T_{{\rm post}}$, $n_{ {\rm post} }$, $\Lambda$, and $R_{gd} = \rho_{{\rm gas} }/\rho_{ {\rm dust} }$ are the post-shock gas pressure, temperature, number density, cooling rate by gas-dust collision per unit volume,
and gas-dust mass ratio, which is typically 100 in the interstellar gas (see \citealt{t05}).
Using the relation between the kinetic energy of the bulk motion in the pre-shock gas and thermal energy in the post-shock gas, we obtain
\[ \frac{  1 }{ 2 } V_{acc}^{2} = \frac{  p_{ {\rm post} }  }{ \rho_{ {\rm post} } (\gamma - 1) } = \frac{  k_{ {\rm bol} } T_{{\rm post}}  }{ m_{ {\rm gas} } ( \gamma - 1  )  },  \]
where $m_{{\rm gas}}$ is the mass of a gas particle. It is straightforward to derive

\[ N_{ {\rm warm} }  \sim n_{  {\rm pre} }  V_{  {\rm acc} }  \tau_{ {\rm cool} } \sim 4.2 \times 10^{20} \left( \frac{ a_{ {\rm dust} }  }{ 0.1 \mu m }  \right)  \left(  \frac{ \rho_{ {\rm mat} }  }{  3.0~ {\rm g ~cm}^{-3} }  \right)  \left( \frac{ R_{gd} }{ 100 }  \right)  \left( \frac{ 6.0 }{ n_{ {\rm post} } /n_{ {\rm pre} } } \right) ~  {\rm cm^{-2}} .\]
We can see that $N_{ {\rm warm} }$ does not depend on the pre-shock gas density.

\begin{figure}
\epsscale{0.80}
\plotone{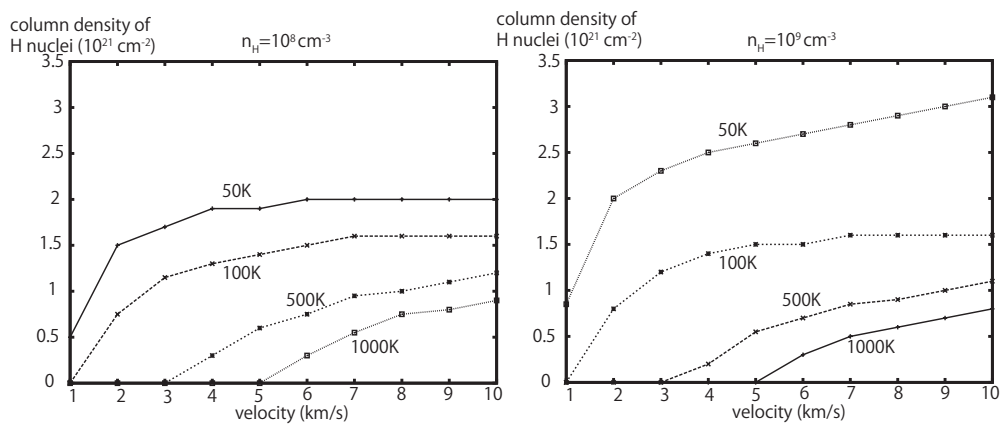}
\caption{Column density of the post-shock gas warmer than 50, 100, 500, and 1000K, 
as a function of the pre-shock velocity, when the pre-shock density is ($n_{\rm H}$) $10^{8}$ cm$^{-3}$ (left panel) and $10^{9}$ cm$^{-3}$ (right panel).}
\label{N_warm}
\end{figure}

Figure \ref{N_warm} shows the column density of warm gas $N_{ {\rm warm} }$
with a temperature higher than 50, 100, 500, and 1000 K, 
as a function of pre-shock velocity for
pre-shock gas densities of $n_{ {\rm H} } = 10^{8}$ and 10$^{9}$ cm$^{-3}$.
As expected from the analytical estimate, $N_{ {\rm warm} }$ is of the order  $10^{21}$ cm$^{-2}$.
The thickness of the gas, warmer than 100 K is
\begin{equation}
\label{L_warm}
L_{ {\rm warm}} \sim 0.06  \left( \frac{ 10^{8} {\rm ~cm^{-3}} }{ n_{\rm pre} } \right) ~ {\rm AU},
\end{equation}
for a pre-shock velocity $\gtrsim$ 2 km s$^{-1}$.
Dependence of $L_{ {\rm warm}}$ on the pre-shock velocity is weaker than that of $N_{ {\rm warm} }$, 
because the post-shock gas density is lower in a lower-velocity shock.
The thickness $L_{ {\rm warm}}$ depends on the pre-shock gas density, because the cooling time $\tau_{ {\rm cool} }$ is shorter at higher densities.

\section{Discussion} 
The radial size of the SO ring around L1527 is about 100 AU. 
\cite{tobin12} estimated the mass of the central
protostar to be $0.19\pm 0.04$ M$_{\odot}$ by fitting the gas velocity at $\lesssim 100$ AU with the Keplerian rotation.
\cite{ohashi14}, on the other hand, found that the power-law index of the rotation profile ($v_{\rm rot}\propto r^{\alpha}$)
is $\alpha \sim -1$ (i.e. indicative of the infall with rotation) at $r\gtrsim 54$ AU, while it is shallower ($\alpha \sim -0.4$)
at inner radius. If the rotation profile at
$\lesssim 54$ AU is due to the Kepler motion, the mass of the central star is $\sim 0.3$ M$_{\odot}$. The star is more massive,
if the Keplerian disk is smaller. Observations with higher angular resolution are desirable to confirm the size of the
Keplerian disk and the mass of the central star. Here, let us assuming that the mass of the central star is $0.2-0.3$ M$_{\odot}$.
Then the free-fall velocity at 100 AU is about 2 km s$^{-1}$.
The gas density in the envelope is not well constrained in the observations, but the density of $\gtrsim 10^{7}$ cm$^{-3}$ is consistent with the model of L1527 by \cite{thc13}.
Our 1D shock model shows that the desorption of SO is possible with $V_{ {\rm acc} } \sim 2$ km s$^{-1}$, when the pre-shock
gas density is $\gtrsim 10^{9}$ cm$^{-3}$.

So far we have assumed that the binding energy of SO is 2600 K, which is for water-dominated ice \citep{gh06}.
The binding energy however, varies with composition and structure (e.g. crystal or amorphous) of the ice mantle.
In \cite{hh93}, the binding energy of SO is estimated to be 2000 K by the summation of van der Waals interaction
of the molecule with the grain surface.
If we adopt this lower value, the gas density required to desorb SO is slightly lower than $10^{9}$ cm$^{-3}$.
The high gas density needed for desorption of SO via the accretion shock is actually consistent with the model by \cite{ssh14};
the P-V diagram of c-C$_3$H$_2$ is reproduced by the accretion of the rotating envelope in the equatorial plane.

The molecules with higher binding energies such as H$_{2}$O and CH$_{3}$OH are difficult to desorb with V$_{ {\rm acc} } \sim 2$ km s$^{-1}$.
In our model, CO$_{2}$, C$_{2}$H$_{2}$, CCS, CH$_{2}$PH, CH$_{3}$CHO, H$_{2}$CN, H$_{2}$CS, H$_{2}$S, NO$_{2}$, and OCN have similar binding energies to SO.
They can be detected in the SO ring, if they are abundant in the ice in the protostellar envelope.

With $V_{ {\rm acc} } = 2$ km s$^{-1}$ and $n_{\rm H} \sim 10^9$ cm$^{-3}$, the column density of warm gas $T\gtrsim$
several tens of K is $\sim 1 \times 10^{21}$ cm$^{-2}$.
According to the chemical models of protostellar cores (e.g. \citealt{whh11,awhgh12}),
SO ice is one of the major carriers of sulfur. If the SO ice abundance in the envelope is $\sim 10^{-7}$,
which is possible, the column density of warm SO is $10^{14}$ cm$^{-2}$.
\cite{ssh14} derived SO temperature $\ge$ 60 K and a column density $\sim 10^{14}$ cm$^{-2}$.
We can conclude that the SO emission around L1527 is consistent with the 1D J-shock.

Although the geometry of the accretion shock might be more complicated than the 1D shock,
the model predicts that the warm SO layer would be geometrically thin (equation (\ref{L_warm})).
When the post-shock gas is cooled, SO will again be absorbed onto grain surfaces in $\sim 10 (10^{9} ~{\rm cm^{-3}}/n_{ {\rm H} } ) $ yr,
if the protostellar heating is not high enough to desorb SO.
Since the absorption timescale is longer than the cooling timescale (equation (\ref{cool_t})),
we also expect a layer of cold SO gas inside the warm SO ring.
The geometrical thickness of the cold SO layer however, depends on the density and velocity structures inside the ring.

\section{SUMMARY}
In this work, we calculate the efficiency of sputtering and thermal desorption of grain surface species in 1D shock with pre-shock gas densities $n_{ {\rm H} } = 10^{6} - 10^{9}$ cm$^{-3}$
and velocities $1 - 10$ km s$^{-1}$.
We compare our results with the recent observation of warm SO emission around the young protostellar core L1527 to investigate if such emission could originate in the shock as the gas accretes on the forming protoplanetary disk.
Our findings are as follows
\begin{itemize}
\item{The thermal desorption is more effective than the sputtering in the accretion shock onto a protoplanetary disk, because of the high gas density.}
\item{We derived the parameter range of gas density and velocity with which volatiles, such as H$_{2}$O and SO, are desorbed to the gas phase.
For example, SO is thermally desorbed in the shock with $V_{ {\rm acc} } \ge 2$ km s$^{-1}$,  if the gas density is higher than $n_{ {\rm H} } = 10^{9}$ cm$^{-3}$,
while the sputtering is negligible at $n_{ {\rm H} } \gtrsim 10^7$ cm$^{-3}$ .}
\item{We show that the column density of warm gas $\sim 100$ K in the post-shock region is $\sim 10^{21}$ cm$^{-2}$.
It only weakly depends on the pre-shock density and velocity, as long as the gas-dust collision is the main cooling mechanism in the post-shock gas.}
\item{The temperature and column density of SO estimated from the observation are consistent with the accretion shock from the envelope onto the forming protoplanetary disk. }

\end{itemize}

\acknowledgments
Numerical computations were in part carried out on the XC30 system at the 
Center for Computational Astrophysics (CfCA) of the National Astronomical 
Observatory of Japan.
This work is supported by Grant-in-aids from the Ministry of Education,
Culture, Sports, Science, and Technology (MEXT) of Japan,
No. 23740154 (T.I.), No. 23540266 and 23103004 (Y.A.).






\appendix

\section{The Mean Velocity of Gas Particles Colliding on a Grain}
\label{mean-vel}
We consider a spherical dust grain with radius $a_{ {\rm dust} }$ moving in the $z$ direction with velocity $V$ relative to gas.
Using the Maxwellian velocity distribution function in the coordinate system of a grain, mean gas velocity can be expressed as 
\[  \int  v  \left(  \frac{  m_{ {\rm gas}  } }{  2 \pi k_{{\rm bol}} T_{{\rm gas}}  }  \right)^{ 3/2 }  {\rm exp} \left( - \frac{ m_{  {\rm gas} } }{   2k_{{\rm bol}} T_{ {\rm gas} }  }  ( v_{x}^{2} + v_{y}^{2}  + ( v_{z} - V )^{2}  )  \right) dv~v~d\theta~v~ {\rm sin} \theta~d\phi , \]
where $\theta$ and $\phi$ are the polar and azimuthal angle in the velocity space ($v_{x}, v_{y}, v_{z}$).
Using the equations of 
$v^{2} = v_{x}^{2}  + v_{y}^{2} + v_{z}^{2} $ and $v_{z} = v  {\rm cos} \theta$, we integrate
the above equation over $\phi$ (0 $\to 2 \pi$) and $\theta$ ( $0 \to \pi$ ). It is straightforward to derive the mean velocity
\[  \int dx~ x^{2}  \left(  \frac{  8 m_{ {\rm gas} }  T_{{\rm gas}}  }{  \pi k_{{\rm bol}} }  \right)^{ 1/2 }  \frac{ 1 }{ 2s } \left(  e^{ -(x-s)^{2} } - e^{  -(x+s)^{2} }   \right) , \]
where $x$ and $s$ are defined as 
 $ x = v/\sqrt{ 2k_{{\rm bol}} T_{{\rm gas}}/m_{ {\rm gas} } } $ and  $ s = V/ \sqrt{ 2k_{{\rm bol}} T_{{\rm gas}}/m_{ {\rm gas} } } $.

\section{Dust Temperature}
\label{d-tem}
We consider a spherical dust grain with radius $a_{ {\rm dust} }$ moving with velocity $v$ relative to the gas of mass $m_p$ and
number density $n_{ {\rm gas} }$.
When the dust collides with a gas particle, the kinetic energy of the gas particle $ \frac{1}{2}m_{p}v^{2} $ is
given to the dust grain. The collision frequency of a dust particle with gas particles is $\pi a_{ {\rm dust} }^{2} v n_{{\rm gas}}$.
The dust cooling rate by thermal radiation per unit surface area is $ \epsilon \sigma_{SB} T^{4}_{ {\rm dust} }$.
We assume that the dust is in thermal equilibrium; the heating rate is equal to the cooling rate
\[  \pi a_{ {\rm dust} }^{2} v  n_{ {\rm gas} } \left(  \frac{1}{2}m_{p}v^{2}  \right)  =   4\pi a_{ {\rm dust} }^{2} \epsilon \sigma_{SB} T^{4}_{ {\rm dust} } .  \]
We adopt $\epsilon \sim 10^{-6} T^{2}$ for a grain with $a_{ {\rm dust} } = 0.1 \mu $m (\citealt{dl84, t05}).
For simplicity, we assume that all the gas particles are hydrogen molecules (H$_{2}$), which means $m_{p} = 3.34 \times 10^{-24}$ g and $n_{ {\rm gas} } = n_{{\rm H_{2} }}$.
It is straightforward to derive 
\[  T_{{\rm dust}} \sim 14 \left(  \frac{ v }{  1~ {\rm km~ s^{-1}} }  \right)^{0.5}  \left(  \frac{  n_{ {\rm H_{2}} } }{ 10^{6} ~ {\rm cm^{-3}}  }     \right)^{1/6}  ~ ~ [K]   . \]
We can see that it is difficult to raise the dust temperature in the shock wave when the number density of gas is low; i.e. $n_{ {\rm H_{2}} } << 10^{6}$ cm$^{-3}$.

\newpage

\begin{table}[htb]
\begin{center}
\begin{tabular}{cc}
\hline
$n_{ {\rm H} }$ (cm$^{-3}$) & $L_{cal}   $(pc)  \\
\hline
$10^{6}$  &  1.0$\times 10^{-3}$    \\
$10^{7}$  &  1.5$\times 10^{-4}$    \\
$10^{8}$  &  2.5$\times 10^{-5}$    \\
$10^{9}$  &  2.0$\times 10^{-6}$    \\
\hline
\end{tabular}
\caption{The size of our numerical domain in the shock models.}
\label{parameter}
\end{center}
\end{table}

\clearpage

\begin{table}[htb]
\begin{center}
\begin{tabular}{|cc|cc|}

\hline
Species & Abundance & Species & Abundance \\
\hline
H$_{2}$ & 0.5 & H$_{2}$S & 1.0$\times$ 10$^{-7}$ \\
He & 0.1 & OCS & 5.0$\times$ 10$^{-8}$ \\
H$_{2}$O & 2.8$\times$ 10$^{-4}$ & Si & 3.6$\times$ 10$^{-8}$ \\
CO & 1.3$\times$ 10$^{-4}$ & PH$_{3}$ & 1.2$\times$ 10$^{-8}$ \\
H & 5.0$\times$ 10$^{-5}$ & S & 5.0$\times$ 10$^{-9}$ \\
N$_{2}$ & 3.7$\times$ 10$^{-5}$ & C$_{2}$H$_{4}$ & 5.0$\times$ 10$^{-9}$ \\
CO$_{2}$ & 3.0$\times$ 10$^{-6}$ & C$_{2}$H$_{5}$OH & 5.0$\times$ 10$^{-9}$ \\
H$_{2}$CO & 2.0$\times$ 10$^{-6}$ & C$_{2}$H$_{6}$ & 5.0$\times$ 10$^{-9}$ \\
O$_{2}$ & 1.0$\times$ 10$^{-6}$ & Fe$^{+}$ & 2.4$\times$ 10$^{-8}$ \\
NH$_{3}$ & 6.0$\times$ 10$^{-7}$ & H$^{+}_{3}$ & 1.0$\times$ 10$^{-9}$ \\
C$_{2}$H$_{2}$ & 5.0$\times$ 10$^{-7}$ & H$^{+}$ & 1.0$\times$ 10$^{-11}$ \\
CH$_{4}$ & 2.0$\times$ 10$^{-7}$ & He$^{+}$ & 2.5$\times$ 10$^{-12}$ \\
CH$_{3}$OH & 2.0$\times$ 10$^{-7}$ & & \\
\hline

\end{tabular}

\caption{Initial molecular abundances relative to hydrogen nuclei}
\label{abund}
\end{center}
\end{table}

\begin{figure}
\epsscale{0.60}
\plotone{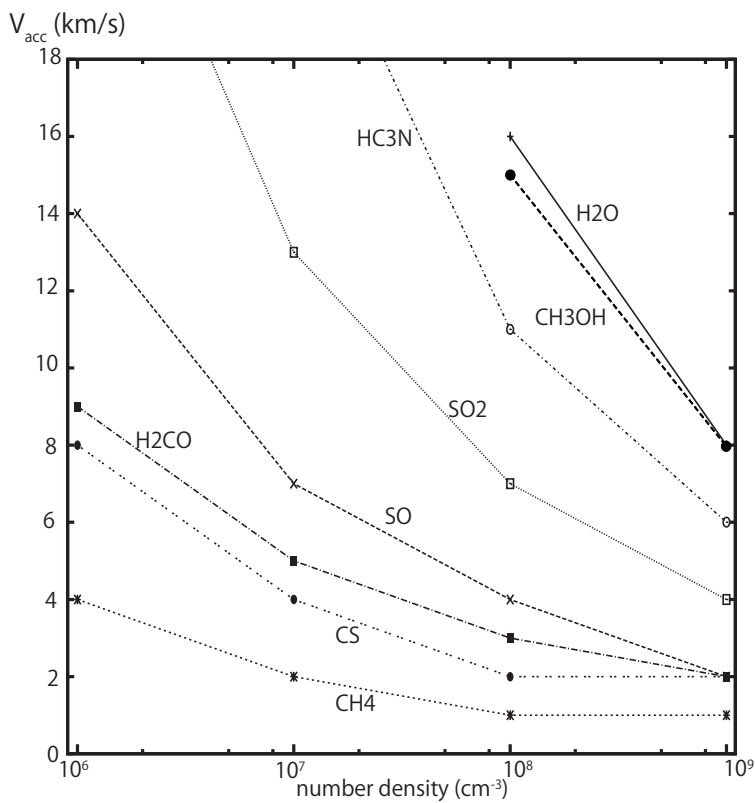}
\caption{The preshock critical velocity above which H$_{2}$O, SO, CO$_{2}$, and CH$_{4}$ are completely desorbed
as a function of the pre-shock density.
The critical velocity of CO$_2$ is almost the same as that of SO, because of their similar binding energies.}
\label{boundary}
\end{figure}

\clearpage

\clearpage


\clearpage



\clearpage




\end{document}